\documentclass[twocolumn,showpacs,preprintnumbers,amsmath,amssymb,superscriptaddress,floatfix,nofootinbib]{revtex4}

\usepackage{graphicx}
\usepackage{amsmath}
\usepackage{amsfonts}
\usepackage{amssymb}

\begin{document}

\title{$\bar{B}^0$, $\bar{B}^0_s$ and $B^-$ decays into $\eta_c$ plus a scalar meson}

\author{Ju-Jun Xie}

\affiliation{Institute of Modern Physics, Chinese Academy of
Sciences, Lanzhou 730000, China}

\author{Gang Li} \email{gli@mail.qfnu.edu.cn}

\affiliation{School of Physics and Engineering, Qufu Normal
University, Shandong 273165, China}

\date{\today}

\begin{abstract}

We investigate the decays of $\bar{B}^0_s$, $\bar{B}^0$ and $B^-$
into $\eta_c$ plus a scalar meson in a theoretical framework by
taking into account the dominant process for the weak decay of
$\bar{B}$ meson into $\eta_c$ and a $q\bar{q}$ pair. After
hadronization of this $q\bar{q}$ component into pairs of
pseudoscalar mesons we obtain certain weights for the pseudoscalar
meson-pseudoscalar meson components. The calculation is based on the
postulation that the scalar mesons $f_0(500)$, $f_0(980)$ and
$a_0(980)$ are dynamically generated states from the pseudoscalar
meson-pseudoscalar meson interactions in $S$-wave. Up to a global
normalization factor, the $\pi \pi$, $K \bar{K}$ and $\pi \eta$
invariant mass distributions for the decays of $\bar B^0_s \to
\eta_c \pi^+ \pi^-$, $\bar{B}^0_s \to \eta_c K^+ K^-$, $\bar B^0 \to
\eta_c \pi^+ \pi^-$, $\bar{B}^0 \to \eta_c K^+ K^-$, $\bar{B}^0 \to
\eta_c \pi^0 \eta$, $B^- \to \eta_c K^0 K^-$ and $B^- \to \eta_c
\pi^- \eta$ are predicted. Comparison is made with the limited
experimental information available and other theoretical
calcualtions. Further comparison of these results with coming LHCb
measurements will be very valuable to make progress in our
understanding of the nature of the low lying scalar mesons,
$f_0(500), f_0(980)$ and $a_0(980)$.
\end{abstract}

\maketitle

\section{Introduction}

In addition to the measurement of the $B^0_s \to J/\psi \pi^+ \pi^-$
decay~\cite{Aaij:2011fx}, the branching fractions ${\rm Br}(B^0_s
\to \eta_c \pi^+ \pi^-) = (1.76 \pm 0.59 \pm 0.12 \pm 0.29) \times
10^{-4}$ and ${\rm Br}(B^0_s \to \eta_c \phi) = (5.01 \pm 0.53 \pm
0.27 \pm 0.63) \times 10^{-4}$ are recently measured by the LHCb
collaboration~\cite{Aaij:2017hfc}. The $f_0(980)$ is produced in the
$\bar B^0_s$ decays into $J/\psi$ and $\pi^+ \pi^-$ and no trace of
the $f_0(500)$ is seen~\cite{Aaij:2011fx}, while in the $\bar B^0
\to J/\psi \pi^+ \pi^-$ decay, the main contribution is from the
$f_0(500)$ with a small fraction for the
$f_0(980)$~\cite{Aaij:2013zpt,Aaij:2014siy}. The new measurement in
Ref.~\cite{Aaij:2017hfc}, suggests also that the $\pi^+ \pi^-$ pair
in $B^0_s \to \eta_c \pi^+ \pi^-$ arises from the contribution of
$f_0(980)$. To understand the new experimental measurements and
search for some hints about involved physics, corresponding
theoretical studies are needed.

Estimations of the branch ratios for some of these decays have been
done by employing the perturbative QCD factorization
approach~\cite{Li:2015tja,Li:2017obb}. Also, in
Ref.~\cite{Ke:2017wni} the decay widths of $B^0_s \to \eta_c
f_0(980)$ and $B^0_s \to \eta_c \phi$ were evaluated in the
light-front quark model. The conclusions of Ref.~\cite{Ke:2017wni}
are that the mostly dominant contribution for the $B^0_s \to \eta_c
\pi^+ \pi^-$ decay is from the $f_0(980)$ and the $f_0(980)$ should
be a $K\bar{K}$ molecule or a tetraquark state, at least its pure
quark-antiquark component is small.

For the $B^0_s \to J/\psi \pi^+ \pi^-$ decay, a simple theoretical
method based on the final state interaction of mesons provided by
the chiral unitary approach has been applied in
Ref.~\cite{Liang:2014tia}, where the theoretical results are in
agreement with the data. The work of Ref.~\cite{Liang:2014tia}
isolates the dominant weak decay mechanism into $J/\psi$ and a
$q\bar{q}$ pair. Then, the $q\bar{q}$ pair is hadronized, and
meson-meson pars are formed with a certain weight. The final state
interaction of the meson-meson components, described in the terms of
chiral unitary theory, gives rise to the $f_0(980)$ and $f_0(500)$
resonances. The approach of Ref.~\cite{Liang:2014tia} was
succesfully extended to study other weak $B$ and $D$ decays in
Refs.~\cite{Bayar:2014qha,Xie:2014tma,Xie:2014gla,Liang:2014ama,Liang:2015qva,Dai:2015bcc,Albaladejo:2016hae,Molina:2016pbg}
(see also Ref.~\cite{Oset:2016lyh} for an extensive review). Other
theoretical work has also been done within the perturbative QCD
approach in Ref.~\cite{Wang:2015uea}. Recently, another approach has
been used in Ref.~\cite{Daub:2015xja} using effective Hamiltonians,
transversity form factors and implementing the meson-meson final
sate interaction. In addition to the $\pi^+ \pi^-$ production, the
$\bar B^0_s$ decay into $J/\psi$ and $K^+ K^-$ is also studied and
compared to experimental measurements in Ref.~\cite{Daub:2015xja}.

Following this line of research, the purpose of this paper is to
investigate the decays of $\bar{B}^0$, $\bar{B}^0_s$ and $B^-$
decays into $\eta_c$ plus a scalar meson. We evaluate the $\pi^+
\pi^-$ and $K^+ K^-$ invariant mass distributions in the $\bar
B^0_s$ decays into $\eta_c \pi^+ \pi^-$ and $\eta_c K^+ K^-$ and the
$K^+ K^-$ and $\pi \eta$ production in the $\bar B^0$ decay into
$\eta_c$ and this pair of mesons. At the same time, we investigate
also the $B^- \to \eta_c K^0 K^-$ and $B^- \to \eta_c \pi^- \eta$
decays. Up to a global factor, one can compare the strength of those
invariant mass distributions.

To end this introduction, we would like to mention that, up to an
arbitrary normalization, one can obtain the invariant mass
distributions and relate the different mass distributions with no
parameters fitted to the data. This is due to the unified picture
that the chiral unitary approach provides for the final state
interaction of mesons. In this sense, predictions on the coming
measurements should be most welcome, and if supported by experiment,
it can give us more information about the nature of these low lying
scalar mesons, $f_0(500)$, $f_0(980)$ and $a_0(980)$, which are
dynamically generated states from the interaction of pseudoscalar
mesons using a meson-meson interaction derived from the chiral
Lagrangians~\cite{Gasser:1983yg,Bernard:1995dp}.

This article is organized as follows. In Sec.~\ref{Sec:Formalism},
we present the theoretical formalism of the decays of $\bar{B}^0$,
$\bar{B}^0_s$ and $B^-$ decays into $\eta_c$ plus a scalar meson,
explaining in detail the hadronization and final state interactions
of the meson-meson pairs. Numerical results and discussions are
presented in Sec.~\ref{Sec:Results}, followed by a summary in the
last section.

\section{Formalism and ingredients} \label{Sec:Formalism}

The leading contributions to the decays of $\bar{B}^0_s$,
$\bar{B}^0$ and $B^-$ into $\eta_c$ plus a scalar meson is the $b
\to c \bar{c} s$ process. In the following we will discuss the
production mechanisms for these decays.

\begin{figure}[htbp]\centering
\includegraphics[scale=0.75]{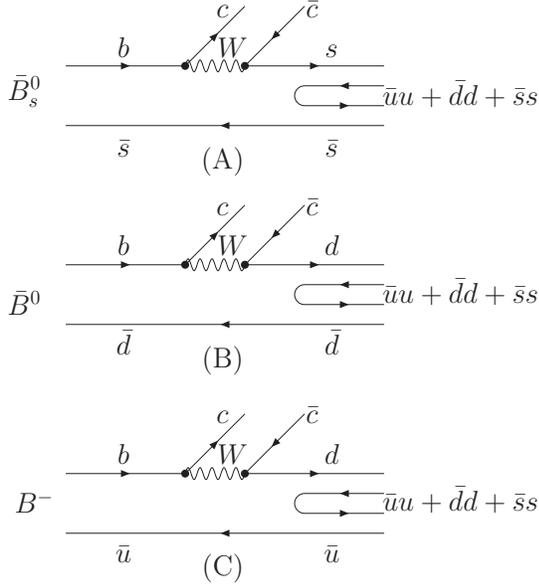}
\caption{Diagrams for the decay of $\bar B^0$, $\bar B^0_s$ and
$B^-$ into $\eta_c$ ($c\bar{c}$) and a primary $q\bar q$ pair,
$s\bar s$ for $\bar B^0_s$ [(A)], $d \bar d$ for $\bar B^0$ [(B)],
and $d \bar{u}$ for $B^-$ [(C)]. The schematic representation of the
hadronization $\bar{q}q \to \bar{q}q(\bar{u}u + \bar{d}d +
\bar{s}s)$ is also shown. \label{fig:feyndiagram}}
\end{figure}

Following Refs.~\cite{Liang:2014tia,Stone:2013eaa}, in
Fig.~\ref{fig:feyndiagram} we show the diagrams at the quark level
that are responsible for the $\bar B^0_s$, $\bar B^0$, and $B^-$
decays into $\eta_c$ and another pair of quarks: $s \bar s$ in the
case of the $\bar B^0_s$ decay [Fig.~\ref{fig:feyndiagram} (A)], $d
\bar d$ in the case of $\bar B^0$ decay [Fig.~\ref{fig:feyndiagram}
(B)], and $d \bar{u}$ for the $B^-$ decay
[Fig.~\ref{fig:feyndiagram} (C)]. The $\bar{B}^0_s$ decay involves
the $V_{cs}$, Cabibbo favored Cabibbo-Kobayashi-Maskawa matrix
element, and the $\bar{B}^0$ and $B^-$ decays involves the $V_{cd}$
Cabibbo suppressed one, which makes the widths large in the
$\bar{B}^0_s$ case compared to the $\bar B^0$ and $B^-$
decays.~\footnote{The use of charge-conjugate modes is implied
throughout this paper.}

In order to produce two mesons the $q \bar q$ pair has to hadronize,
which one can implement adding an extra $\bar q q $ pair with the
quantum numbers of the vacuum, $\bar u u+ \bar d d+ \bar s s$, see
also in Fig.~\ref{fig:feyndiagram}. Next step corresponds to writing
the $q\bar{q} (\bar u u+ \bar d d+ \bar s s)$ combination in terms
of pairs of mesons. Following the work of Ref.~\cite{Liang:2014tia}
we obtain
\begin{eqnarray}
d\bar d (\bar u u +\bar d d +\bar s s) && \equiv   \pi^- \pi^+
+\frac{1}{2}\pi^0 \pi^0
+\frac{1}{3}\eta \eta \nonumber \\
&& -\frac{2}{\sqrt{6}}\pi^0 \eta +\bar K^0 K^0 , \label{eq:ddbarhad} \\
s\bar s (\bar u u + \bar d d +\bar s s) && \equiv K^- K^+ + \bar K^0
K^0 +\frac{1}{3}\eta \eta, \label{eq:ssbarhad} \\
d\bar u (\bar u u +\bar d d +\bar s s) && \equiv
\frac{2}{\sqrt{3}}\pi^- \eta + K^0 K^- , \label{eq:dubarhad}
\end{eqnarray}
where the $\eta'$ terms have been neglected because the $\eta'$ has
large mass and has very small effect here.

\begin{figure*}[htbp]\centering
\includegraphics[scale=0.9]{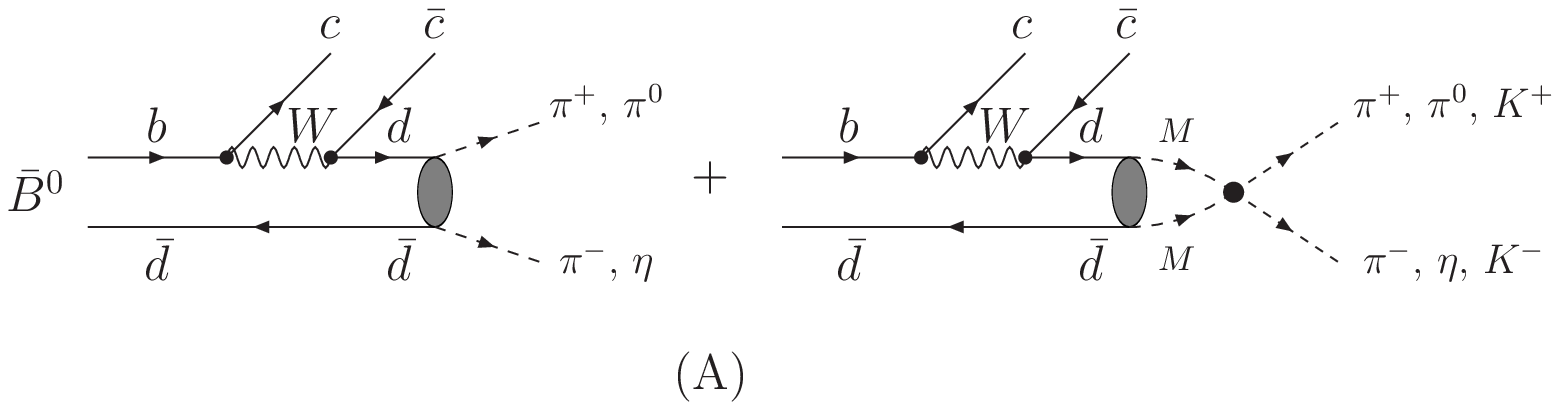}
\includegraphics[scale=0.9]{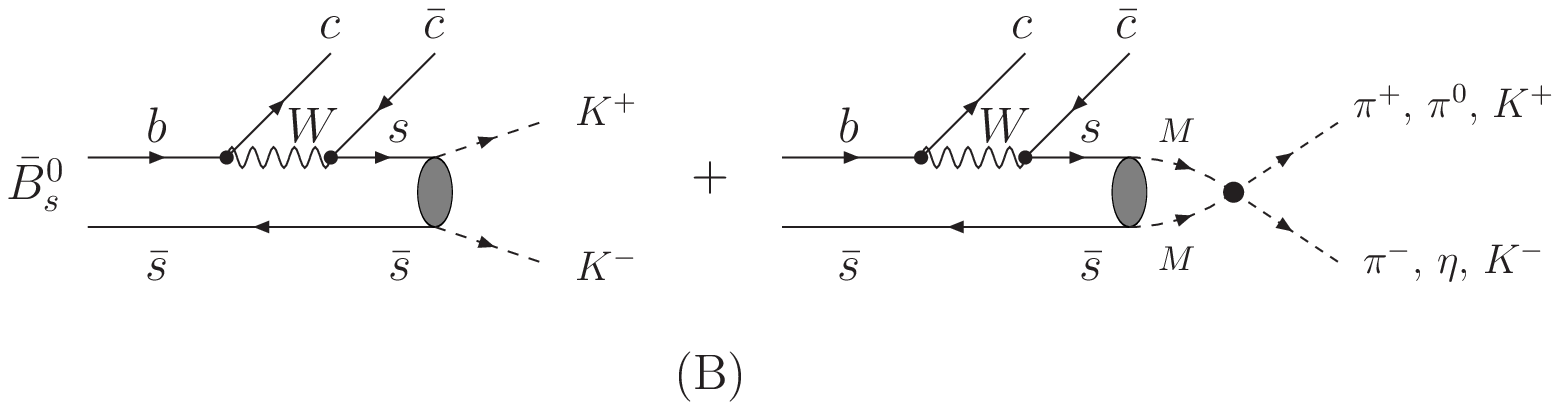}
\includegraphics[scale=0.9]{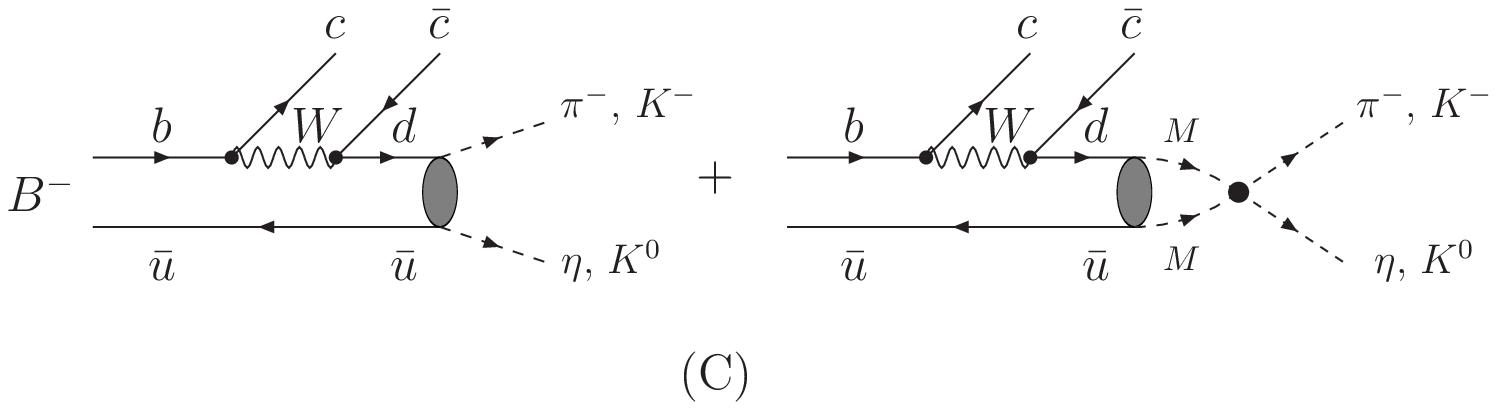}
\caption{Diagrammatic representations of the production of $\pi^+
\pi^-$, $\pi^0 \eta$, $K^+ K^-$, $\pi^- \eta$, and $K^-K^0$ via
direct plus re-scattering mechanisms in $\bar B^0$ (A), $\bar B
^0_s$ (B) and $B^-$ (C) decays. \label{fig:fsi}}
\end{figure*}

After the production of a meson-meson (MM) pair, the final state
interaction between the meson and meson takes place, which can be
parameterized by the re-scattering shown in Fig.~\ref{fig:fsi} at
the hadronic level. Since we consider only the $S$-wave interaction
between the pseudo-scalar meson and pseudo-scalar meson, we will
have the contributions from only the scalar mesons. In
Fig.~\ref{fig:fsi}, we also show the tree level diagrams for the
$\pi \pi$, $K \bar{K}$ and $\pi \eta$ production.

The decay amplitudes for a final production of the different meson
pairs are given by~\cite{Oller:1997yg}
\begin{eqnarray}
&& T(\bar B^0_s \to \eta_c \pi^+ \pi^-) = V_P V_{cs} ( G_{K^+
K^-} t_{K^+ K^- \to \pi^+ \pi^-}  \nonumber \\
&&  +  G_{K^0 \bar K^0} t_{K^0 \bar K^0 \to \pi^+ \pi^-} +
\frac{1}{3} G_{\eta \eta} t_{\eta
\eta \to \pi^+ \pi^-} ), \label{eq:tBspipi} \\
&& T(\bar B^0_s \to \eta_c K^+ K^-)  = V_P V_{cs} ( 1+ G_{K^+ K^-} t_{K^+ K^- \to K^+ K^-} \nonumber \\
&& +  G_{K^0 \bar K^0} t_{K^0 \bar K^0 \to K^+ K^-}  + \frac{1}{3}
G_{\eta \eta} t_{\eta \eta \to K^+ K^-} ), \label{eq:tBskaka} \\
&& T(\bar B^0 \to \eta_c \pi^+ \pi^-) = V_P V_{cd} (
1+G_{\pi^+ \pi^-} t_{\pi^+ \pi^- \to \pi^+ \pi^-} \nonumber \\
&& + \frac{1}{2} G_{\pi^0 \pi^0} t_{\pi^0 \pi^0 \to \pi^+ \pi^-}  +
G_{K^0 \bar K^0} t_{K^0 \bar K^0 \to \pi^+ \pi^-} \nonumber \\
&& + \frac{1}{3} G_{\eta \eta} t_{\eta \eta \to \pi^+ \pi^-}
),\label{eq:tBzeropipi}
\\
&& T(\bar B^0 \to \eta_c \pi^0 \eta)  = V_P V_{cd}
(-\frac{2}{\sqrt{6}}-\frac{2}{\sqrt{6}} G_{\pi^0 \eta} t_{\pi^0 \eta
\to \pi^0 \eta} \nonumber \\
&&  + G_{K^0 \bar K^0} t_{K^0 \bar K^0\to\pi^0 \eta} )
\label{eq:tBzeropieta}, \\
&& T(\bar B^0 \to \eta_c K^+ K^-) = V_P V_{cd} ( G_{\pi^+ \pi^-}
t_{\pi^+ \pi^- \to K^+ K^-} \nonumber \\
&& + \frac{1}{2} G_{\pi^0 \pi^0} t_{\pi^0 \pi^0 \to K^+ K^-} +
\frac{1}{3} G_{\eta \eta} t_{\eta \eta \to K^+ K^-} \nonumber \\
&& -\frac{2}{\sqrt{6}} G_{\pi^0 \eta} t_{\pi^0 \eta \to K^+ K^-}
+G_{K^0 \bar K^0} t_{K^0 \bar K^0\to K^+ K^-}
),\label{eq:tBzerokaka} \\
&& T(B^- \to \eta_c \pi^- \eta) = V_P V_{cd} (
\frac{2}{\sqrt{3}}+\frac{2}{\sqrt{3}} G_{\pi^- \eta} t_{\pi^- \eta
\to \pi^- \eta} \nonumber \\
&& +G_{K^0 K^-} t_{K^0 K^- \to \pi^- \eta} ),
\label{eq:tBminuspieta} \\
&& T(B^- \to \eta_c K^0 K^-) = V_P V_{cd} (1+G_{K^0 K^-} t_{K^0
K^- \to K^0 K^-} \nonumber \\
&& + \frac{2}{\sqrt{3}} G_{\pi^- \eta} t_{\pi^- \eta \to K^0 K^-} ).
\label{eq:tBminuskaka}
\end{eqnarray}
where $V_P$ is the production vertex which contains all dynamical
factors common to all the above seven decays. We shall assume $V_P$
as constant and fit it to the experimental date. The $G_{\rm MM}$
are the loop functions of two meson propagators. The $t_{{\rm MM}\to
{\rm MM}}$ are the scattering matrices and they are calculated in
Ref.~\cite{Liang:2014tia} following Ref.~\cite{Oller:1997ti}. Note
that we can easily obatin $t_{\pi^- \eta \to \pi^- \eta}$, $t_{K^0
K^- \to K^0 K^-}$ and $t_{\pi^- \eta \to K^0 K^-}$ using isospin
symmetry,
\begin{eqnarray}
t_{\pi^- \eta \to \pi^- \eta} &=& t_{\pi^0 \eta \to \pi^0 \eta}, \label{eq:tpieta2pieta}\\
t_{K^0 K^- \to \pi^- \eta} &=& -\sqrt{2}~ t_{K^0 \bar K^0 \to \pi^0 \eta}, \label{eq:tK0K2pieta}\\
t_{K^0 K^- \to K^0 K^-} &=& \!\!  t_{K^+ K^- \to K^+ K^-} \! - \!
t_{K^+ K^- \to K^0 \bar K^0}.\label{eq:tK0K2K0K}
\end{eqnarray}

With all the ingredients obtained in the previous section, one can
write down the invariant mass distributions for those decays as
\begin{equation} \label{eq:dGamma}
\frac{d \Gamma}{d M_{\rm inv}} = \frac{1}{(2\pi)^3}\frac{1}{4M_{\bar
B_j}^2} p_{\eta_c} \tilde{p}_{\rm M} \sum \left|T \right|^2,
\end{equation}
where $M_{\bar{B}_j}$ is the mass of $\bar{B}^0$, $\bar{B}^0_s$, or
$B^-$, while $M_{\rm inv}$ is the invariant mass of the final ${\rm
MM}$ pair. The $p_{\eta_c}$ is the $\eta_c$ momentum in the rest
frame of $\bar{B}_j$ and $\tilde{p}_{M}$ is the momentum of one
pseudo-scalar meson in the rest frame of ${\rm MM}$ pair.

\section{Numerical Results and discussions} \label{Sec:Results}

Same to the $\bar{B}^0_s \to J/\psi \pi^+ \pi^-$
decay~\cite{Liang:2014tia}, the $\bar{B}^0_s \to \eta_c \pi^+ \pi^-$
decay is also dominant by $f_0(980)$. In Ref.~\cite{Li:2015tja}, the
fraction for the $f_0(980)$ contribution in the $\bar{B}^0_s \to
\eta_c \pi^+ \pi^-$ decay is around $70\%$. Thus, we
assume~\footnote{The experimental result for the fraction of the
$f_0(980)$ contribution in the $\bar{B}^0_s \to J/\psi \pi^+ \pi^-$
is $(65.0-94.5)\%$ (see Table X of Ref.~\cite{Aaij:2014emv}).}
\begin{eqnarray}
\frac{{\rm Br} [\bar{B}^0_s \to \eta_c f_0(980) \to \eta_c \pi^+
\pi^-]}{{\rm Br}(\bar{B}^0_s \to \eta_c \pi^+ \pi^-)} = (80 \pm
10)\%, \nonumber
\end{eqnarray}
from where we get ${\rm Br} [\bar{B}^0_s \to \eta_c f_0(980) \to
\eta_c \pi^+ \pi^-] = (1.41 \pm 0.56) \times 10^{-4}$, where we have
added in quadrature the three sets of errors quoted in
Ref.~\cite{Aaij:2017hfc}.

On the other hand, if we integrated the Eq.~\eqref{eq:dGamma}, up to
one free parameter $V_P$, we can extract the contribution from
$f_0(980)$ for the decay of $\bar{B}^0_s \to \eta_c \pi^+ \pi^-$,
since, in our production mechanism, the main contribution for this
decay is $f_0(980)$. Then, one can determine $V_P$. With $V_{cs} =
0.97427$, we get
\begin{eqnarray}
V_P = (3.44 \pm 0.68) \times 10^{-6}.  \label{eq:vp}
\end{eqnarray}

\begin{figure}[thb]\centering
\includegraphics[scale=0.45]{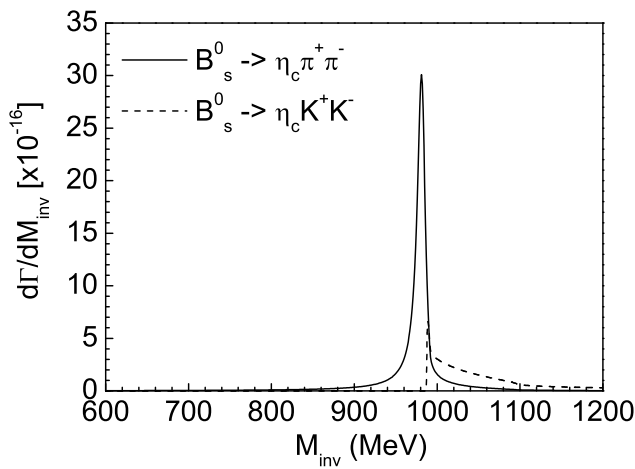}
\caption{$\pi^+ \pi^-$ and $K^+ K^-$ invariant mass distributions
for $\bar B_s^0 \to \eta_c \pi^+ \pi^-$ and $\eta_c K^+ K^-$.
\label{fig:Bsimd}}
\end{figure}

\begin{figure}[thb]\centering
\includegraphics[scale=0.45]{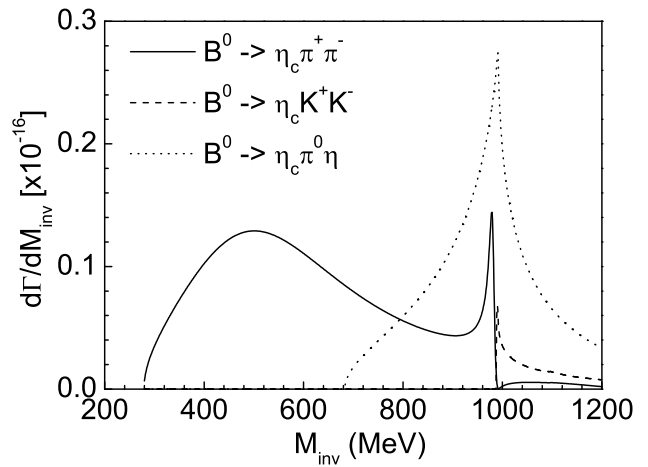}
\caption{$\pi^+ \pi^-$, $\pi^0 \eta$, $K^+ K^-$ invariant mass
distributions for $\bar B^0 \to \eta_c \pi^+ \pi^-$, $\eta_c K^+
K^-$, $\eta_c \pi^0 \eta$. \label{fig:Bzeroimd}}
\end{figure}

\begin{figure}[thb]\centering
\includegraphics[scale=0.45]{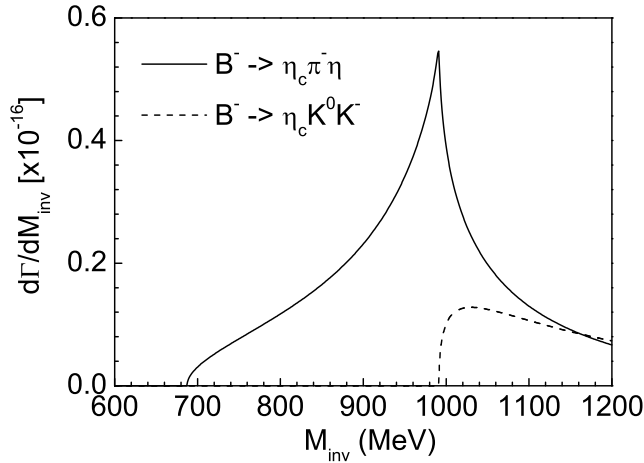}
\caption{$\pi^- \eta$ and $K^0 K^-$ invariant mass distributions for
$B^- \to \eta_c \pi^- \eta$ and $B^- \to \eta_c K^0 K^-$.
\label{fig:Bminusimd}}
\end{figure}

Our theoretical results with $V_{cd} = -0.22534$ and $V_P = 3.44
\times 10^{-6}$ are summarized in Figs.~\ref{fig:Bsimd},
\ref{fig:Bzeroimd}, and \ref{fig:Bminusimd}. In Fig.~\ref{fig:Bsimd}
we show the $\pi^+ \pi^-$ and $K^+ K^-$ invariant mass distributions
for the $\bar B^0_s \to \eta_c \pi^+ \pi^-$ and $\bar B^0_s \to
\eta_c K^+ K^-$, respectively. As one can see, the $f_0(980)$
production is clearly dominant while there is no evident signal for
the $f_0(500)$. For the $\bar B^0_s \to \eta_c K^+ K^-$ decay, the
$K^+ K^-$ distribution gets maximum strength just above the $K^+
K^-$ threshold and then falls down gradually. This is due to the
effect of the $f_0(980)$ resonance below the $K\bar{K}$
threshold.~\footnote{The pole position for $f_0(980)$ is obtained
as: $\sqrt{s_R} = 981.5 -i5.5~ ({\rm MeV})$.} Starting from the
dominant weak decay process we have $\eta_c$ and $s\bar{s}$
production in the $\bar{B}^0_s$ decay. Because $s \bar s$ pair has
isospin zero, and the strong interaction hadronization conserves it.
Even the $K^+ K^-$ system could be $I=0$ or $1$, the process of
formation guarantees that this is an $I=0$ state and the shape of
the $K^+ K^-$ distribution is due to the $f_0(980)$ with $I = 0$.

The strength for the $K^+ K^-$ distribution is small compared to the
one of the $f_0(980)$ at its peak for the $\pi^+ \pi^-$
distribution, but the integrated strength over the invariant mass of
$K^+ K^-$ is of the same order of magnitude as that for the strength
below the peak of the $f_0(980)$ going to $\pi^+ \pi^-$. On the
other hand, we should mention that we are calculating only the
$S$-wave contribution of the $K^+ K^-$ distribution, hence,
contributions from higher waves, such as $\phi$ ($P$-wave),
$f'_2(1525)$ ($D$-wave) etc, are not included. It is interesting to
compare this with experiment. First by integrating the strength of
the $K^+ K^-$ distribution over its invariant mass, up to $M_{\rm
inv}(K^+ K^-) = 1200$ MeV,~\footnote{We should mention that the
chiral unitary approach that we use only makes reliable predictions
up to 1200 MeV~\cite{Oller:1997ti}. One should not use the model for
higher invariant masses. With this perspective we will have to admit
uncertainties in the mass distributions, particularly at invariant
masses higher than 1200
MeV~\cite{Liang:2016hmr,Debastiani:2016ayp,Guo:2016zep}.} we find a
ratio
\begin{equation}\label{eq:ratio-Bskaka2pipi}
\frac{{\rm Br} [ \bar B^0_s \to \eta_c f_0(980) \to \eta_c K^+
K^-]}{{\rm Br} [\bar B^0_s \to \eta_c f_0(980) \to \eta_c \pi^+
\pi^- ]} = 0.4.
\end{equation}

Secondly, if we stick to a band of energies around the $\phi$ meson
peak, $990<M_{\rm inv}(K^+K^-) <1050$ MeV, as done in
Ref.~\cite{Aaij:2013orb} for the $\bar{B}^0_s \to J/\psi K^+K^-$, we
get the $S$-wave fraction
\begin{equation}\label{eq:ratio-BsKK2Bsphi}
\frac{{\rm Br}[\bar B^0_s \to \eta_c K^+ K^-] (S{\rm -wave})}{{\rm
Br} [\bar B^0_s \to \eta_c \phi \to \eta_c K^+ K^- ]}= (13 \pm 6)
\times 10^{-2} ,
\end{equation}
where ${\rm Br} [\bar B^0_s \to \eta_c \phi] = (5.01 \pm 0.87)
\times 10^{-4}$~\cite{Aaij:2017hfc} and the branching fraction of
$0.489$ for $\phi$ decay into $K^+ K^-$ has been
taken~\cite{Patrignani:2016xqp}. This value, one of our model
predictions, could be tested by future experiment.

We come back now to the decays of the $\bar B^0$. In
Fig.~\ref{fig:Bzeroimd} we show the theoretical results for the
$\pi^+ \pi^-$, $K^+ K^-$ and $\pi^0 \eta$, invariant mass
distributions for $\bar B^0 \to \eta_c \pi^+ \pi^-$, $\eta_c K^+
K^-$, and $\eta_c \pi^0 \eta$. In the $\bar B^0$ decays, we had the
hadronization of a $d \bar d$ pair, which contains $I=0$ and $1$.
But, the $\pi^+ \pi^-$ in $S$-wave can only be in $I=0$, hence the
peaks for the $\pi^+ \pi^-$ distribution due to the $f_0(500)$ and
$f_0(980)$ excitation. It is expected that the $\rho^0$ contribution
peaks around 770 MeV, and has larger strength than the $f_0(500)$
contribution, but at invariant masses around 500 MeV and bellow, the
strength of the $f_0(500)$ dominates the one of the $\rho^0$ meson.
For the $K^+ K^-$ production in the $\bar B^0$ decay, we have
considered both the $I=0$ [$f_0(980)$] and $I=1$ [$a_0(980)$]
contribution.

One can see that, from Fig.~\ref{fig:Bzeroimd}, the strength of the
$f_0(980)$ excitation is very small compared to that of the
$f_0(500)$ (the broad peak to the left). Note that because of the
experimental resolution the $f_0(980)$ peak would not appears so
narrow in the experiments. As done in
Refs.~\cite{Liang:2014tia,Xie:2014tma}, we can extract the
$f_0(500)$ contribution to the branching ratio by assuming a smooth
background below the $f_0(980)$ peak, we find
\begin{eqnarray}
{\rm Br} [\bar{B}^0 \to \eta_c f_0(500) \to \eta_c \pi^+ \pi^-] \!\!
= \!\! (1.2 \pm 0.5) \times 10^{-5},
\end{eqnarray}
with error from the uncertainty of $V_P$ shown in Eq.~\eqref{eq:vp}.
Then we find a ratio, $R$
\begin{eqnarray}
R \! \! = \!\! \frac{{\rm Br} [ \bar B^0 \to \eta_c f_0(500) \to
\eta_c \pi^+ \pi^-]}{{\rm Br} [\bar B^0_s \to \eta_c f_0(980) \to
\eta_c \pi^+ \pi^- ]} \!\! = \!\! ( 9 \pm 5) \times 10^{-2},
\end{eqnarray}
which is consistent with the ones obtained in
Ref.~\cite{Li:2015tja}: $R = (3 \sim 8) \times 10^{-2}$ in
Breit-Wigner model and $R = (4 \sim 12) \times 10^{-2}$ in Bugg
model.~\footnote{See details for the definitions of Breit-Wigner and
Bugg models in Ref.~\cite{Li:2015tja}.} However, the branch ratio,
${\rm Br} [\bar{B}^0 \to \eta_c f_0(500) \to \eta_c \pi^+ \pi^-]$
obtained here, is much larger than the one obtained in
Ref.~\cite{Li:2015tja} with the perturbative QCD factorization
approach. We hope the future experimental measurements can clarify
this issue.

In Fig.~\ref{fig:Bzeroimd}, the $\pi^0 \eta$ invariant mass
distribution has a sizeable strength, bigger than that for the
$\pi^+ \pi^-$ and $K^+ K^-$. As one can see, we get the typical cusp
structure of the $a_0(980)$. This prediction is tied exclusively to
the weights of the starting meson meson channels in
Eq.~(\ref{eq:ddbarhad}) and the final state interaction in Eqs.~(\ref{eq:tBzeropipi}),
\eqref{eq:tBzeropieta}, and \eqref{eq:tBzerokaka}. Hence, this is a
prediction of this approach, not tied to any experimental input.

Next, we show the results for $B^-$ decay in
Fig.~\ref{fig:Bminusimd}, where the strength for the $\pi^- \eta$
invariant mass distribution is two times as big as the one of $\bar
B^0 \to \eta_c \pi^0 \eta$ shown in Fig.~\ref{fig:Bzeroimd}. For the
$K^0 K^-$ mass distribution we see that the position of the peak has
moved to higher invariant masses compared to the $K^+ K^-$ invariant
mass spectrum of the $\bar B^0 \to \eta_c K^+ K^-$ or $\bar B_s^0
\to \eta_c K^+ K^-$ decays. In fact, the $K^0 K^-$ invariant mass
distribution in the $B^-$ decay due to the $a_0(980)$, which is seen
in the figures, is much wider than that of the $f_0(980)$. It would
be most instructive to see all these features in future experiments.

\section{Summary}

We have performed a study of the $\pi \pi$, $\pi \eta$ and $K
\bar{K}$ invariant mass distributions for $\bar B^0_s \to \eta_c
\pi^+ \pi^-$, $\bar B^0_s \to \eta_c K^+ K^-$, $\bar B^0 \to \eta_c
\pi^+ \pi^-$, $\bar B^0 \to \eta_c K^+ K^-$, $\bar B^0 \to \eta_c
\pi^0 \eta$, $B^- \to \eta_c \pi^- \eta$, and $B^- \to \eta_c K^0
K^-$. We take the dominant mechanism for the weak decay of the $B$
meson, going to $\eta_c$ and a $q\bar{q}$ pair that, upon
hadronization, leads to $\pi \pi$, $\pi \eta$, and $K\bar{K}$ in the
final state, and this interaction is basically mediated by the
scalar mesons, $f_0(500)$, $f_0(980)$, and $a_0(980)$.

Up to a global factor,\footnote{The model relies on the constancy of
the $V_P$ factor which contains the weak amplitudes and the
hadronization procedure. The only thing demanded is that this factor
is smooth and practically constant as a function of the invariant
masses in the limited range where the predictions are made (see more
details in Refs. \cite{Liang:2014tia,Sekihara:2015iha}.} which is
determined to the experimental measurement, we can compare the
strength of the $\pi \pi$, $\pi \eta$ and $K \bar{K}$ invariant mass
distributions. For the $\bar B^0_s \to \eta_c K^+ K^-$, only the
$f_0(980)$ resonance contributes to the $K^+K^-$ mass distribution,
but in the case of the $\bar B^0 \to \eta_c K^+ K^-$, both the
$f_0(980)$ and $a_0(980)$ resonances contribute to its strength. The
strength of the $K \bar{K}$ invariant mass distribution in the
$\bar{B}^0_s$ decay is much larger than the one in $\bar{B}^0$
decay, which is because the $\bar{B}^0_s$ decay is Cabibbo favored
process, while the $\bar{B}^0$ decay is the Cabibbo suppressed
process. In the case of the $\bar B^0 \to \eta_c \pi^0 \eta$, one
finds a cusp structure for the $a_0(980)$ and its strength is much
larger than the one for the $\bar B^0 \to \eta_c \pi^+ \pi^-$ decay
around the $f_0(980)$ peak.

Our theoretical results shown here are predictions for ongoing
experiments at LHCb, and comparison of the observed results with our
predictions will be most useful to make progress in our
understanding of the meson-meson interaction and the nature of the
low lying scalar mesons.

\section*{Acknowledgments}

This work is also partly supported by the National Natural Science
Foundation of China under Grant Nos. 11475227, 11675091, and 11735003 and the
Youth Innovation Promotion Association CAS (No. 2016367).

\bibliographystyle{plain}

\end{document}